# DYNAMIC SELECTION OF SYMMETRIC KEY CRYPTOGRAPHIC ALGORITHMS FOR SECURING DATA BASED ON VARIOUS PARAMETERS

Ranjeet Masram[1], Vivek Shahare[1], Jibi Abraham[1], Rajni Moona[1], Pradeep Sinha[2], Gaur Sunder[2], Prashant Bendale[2] and Sayali Pophalkar[2]

[1] Department of Computer Engineering and Information Technology, COEP, India
masram.ranjeet@gmail.com, vivek.shahare27@gmail.com,
ja.comp@coep.ac.in, rajnimoona@yahoo.com
[2] C-DAC Pune, India
psinha@cdac.in, gaurs@cdac.in, prashantb@cdac.in, psayali@cdac.in

## ABSTRACT

*Most of the information is in the form of electronic data. A lot of electronic data exchanged takes place through computer applications. Therefore information exchange through these applications needs to be secure. Different cryptographic algorithms are usually used to address these security concerns. However, along with security there are other factors that need to be considered for practical implementation of different cryptographic algorithms like implementation cost and performance. This paper provides comparative analysis of time taken for encryption by seven symmetric key cryptographic algorithms (AES, DES, Triple DES, RC2, Skipjack, Blowfish and RC4) with variation of parameters like different data types, data density, data size and key sizes.*

## 1. INTRODUCTION

Cryptography is a science of "secret messages" that is being used by man from thousands of years [9]. In cryptography original message is basically encoded in some non readable format. This process is called encryption. The only person who knows how to decode the message can get the original information. This process is called decryption. On the basis of key used, cipher algorithms are classified as asymmetric key algorithms, in which encryption and decryption is done by two different keys and symmetric key algorithms, where the same key is used for encryption and decryption [8]. On the basis of the input data, cipher algorithms are classified as block ciphers, in which the size of the block is of fixed size for encryption and stream ciphers in which a continuous stream is passed for encryption and decryption [9].

A data file formats represents the standard for encoding the information to be stored in computer file. There are file formats like textual, image, audio and video data file formats. Textual data formats are ANSII, UNICODE (16 & 32 bit little and big Endian and UTF-8). ANSII is encoding scheme for 128 characters basically made for English alphabets. It contains alphabets a-z and A-Z, numbers 0-9 and some special characters. In Unicode standard unique numbers are provided for every character independent of platform. Image file formats are JPEG, TIFF, BMP, GIF and PNG [10]. JPEG is image file format for digital images that uses lossy compression method. TIFF and BMP are image file format that are used to store images of raster graphics. GIF image is similar to image format of bitmap images. GIF uses LZW (Lempel-Ziv-Welch) technique of compression and for each image it can support up to 8 bits/pixel. PNG is alternative to GIF image file format and allows more compression than GIF. Audio file formats are WAV, AIFF, M4A, MP3 and WMA. WAV and AIFF are usually uncompressed audio file format. M4A (audio) uses Apple Lossless compression format but often it is compressed with Advance audio coding (lossy). MP3 and WMA are lossy compression audio formats. Video file formats are AVI, M4V, MPEG and WMV etc. AVI format contains the data (audio and video data) file container; which allows audio-with-video playback synchronously. M4V and MP4 are very similar format, but M4v can be protected by Digital Rights Management copy protection. MPEG contains compressed audio and visual digital data. WMV is compressed video format developed by Microsoft.

Density of data represents the amount of different information present in the data file [10]. File is said to be dense file if file size is less and content is more. For example if there are two file X and Y both containing 2000 words and having sizes 50kb and 200kb respectively, then file X is denser. The more the information, the dense is the data and lesser the information, sparse is the data. Sparse file is a file that contains most of the empty spaces and attempts to use the computer space more effectively.

Data size is space occupied by a file on a disk. Audio, video takes more space on disk than textual files as they contain multimedia information. Key size in cryptography represents the size of key file in bits. For example AES is having key sizes 128, 192 and 256 bits.

The main objective of this paper is to analyze time taken for encryption by various cryptographic algorithms for parameters like data type, data size, data density and key size.

## 2. CRYPTOGRAPHIC ALGORITHMS

This section provides information about the various symmetric key cryptographic algorithms to be analyzed for performance evaluation, to select the best algorithm with appropriate parameter suitable to provide security for data. The various features of the cryptographic algorithm are listed in Table 1.

Table 1. Cryptographic Algorithms Information.

| Algorithm Name | Structure | Cipher Type | Rounds | Key Size(In bits) |
|---|---|---|---|---|
| AES | Substitution-permutation network | Block | 10, 12, 14 | 128, 192, 256 |
| DES | Balanced Feistel network | Block | 16 | 56 |
| Triple DES | Feistel network | Block | 48 | 112, 168 |
| RC2 | Source-heavy Feistel network | Block | 18 | 40 to 1024 |
| Blowfish | Feistel network | Block | 16 | 32 to 448 |
| Skipjack | Unbalanced Feistel network | Block | 32 | 80 |
| RC4 | ---- | Stream | 256 | 40 to 2048 |

## 3. RELATED WORK

This section provides the information and results which are obtained from the numerous sources. Cryptographic algorithms have been compared with each other for performance evaluation on basis of throughput, CPU Memory utilization, energy consumption, attacks, Encryption time, Decryption time etc.

In [3] the author compared AES and RC4 algorithm and the performance metrics were encryption throughput, CPU work load, memory utilization, and key size variation and encryption and decryption time. Results show that the RC4 is fast and energy saving for encryption and decryption. RC4 proved to be better than AES for larger size data. In [2] author compared AES and DES algorithms on image file, MATLAB software platform was used for implementation of these two cipher algorithms. AES took less encryption and decryption time than DES. In [4] the author compared cipher algorithms (AES, DES, Blowfish) for different cipher block modes (ECB, CBC, CFB, OFB) on different file sizes varying from 3kb to 203kb. Blowfish algorithm yield better performance for all block cipher modes that were tested and OFB block mode gives better performance than other block modes. In [7] the author talks about comparison between three algorithms (DES, Triple DES, Blowfish) on processing time. They found, that the key generation time for all these three algorithms is almost same but there is a difference in time taken by CPU for encryption. On SunOS platform Blowfish seem to be fastest, followed by DES and Triple DES respectively. They analyzed CPU execution time for generating the secret key, encryption and decryption time on 10MB file. In [6] the author compared cipher algorithms (AES, DES, 3-DES and Blowfish) for varying file size and compared the encryption time on two different machines Pentium-4, 2.4 GHz and Pentium-II 266 MHz in EBC and CFB Mode. The author concluded that Blowfish is fastest followed by DES and Triple DES and CFB takes more time than ECB cipher block mode.

## 4. PROPOSED WORK

From the related works, it is realized that none of the work did a very detailed analysis of the performance of various symmetric algorithms on various parameters on different type of files, especially the files which are used for medical health related data.

The main objective of this paper is to analyze the time taken for encryption by various cryptographic algorithms for parameters like data type, data size, data density and key size in order to select the most suitable cryptographic algorithm for encryption.

## 5. EXPERIMENTAL SETUP AND TESTING

The execution results are taken on machine having Intel® Core™ i7-2600 (3.40 GHz) processor with Intel® Q65 Express 4 GB 1333 MHz DDR3 (RAM) and Ubuntu 12.04 LTS operating System. The java platform (openjdk1.6.0_14) is used for implementation. JCA (Java Cryptography Architecture) and JCE (Java Cryptography Extension) are used for cipher algorithm implementation. The JCA is a major platform that contains "provider" architecture and the set of APIs for encryption (symmetric ciphers, asymmetric ciphers, block ciphers, stream ciphers), message digests (hash), digital signatures, certificates and certificate validation, key generation and secure random number generation. Here we have used sun and Bouncy Castel provider for implementing cryptographic algorithms.

The brief analysis of different symmetric key cryptographic algorithm for various parameters is as follows:

**Case Study 1: Files with different Data types.**

This case study has taken to check whether the encryption has dependency on type of data. Different data type files like audio, image, textual and video of nearly 50MB in size are chosen and encryption time of different cipher algorithms is calculated for these data types. For all executions of a specific cipher algorithm, varying parameter is data type and constant parameters are key size and block cipher mode. Key size and block mode are at kept at bare minimal parameters. The key size of AES, DES, 3-DES, RC2, Blowfish, Skipjack, and RC4 are kept at minimum values as 128, 56, 112, 40, 32, 80 and 40 bits respectively. Block cipher mode used is ECB with PKCS#5 padding scheme. Fig. 1 shows the execution time of the algorithms for different data type files.

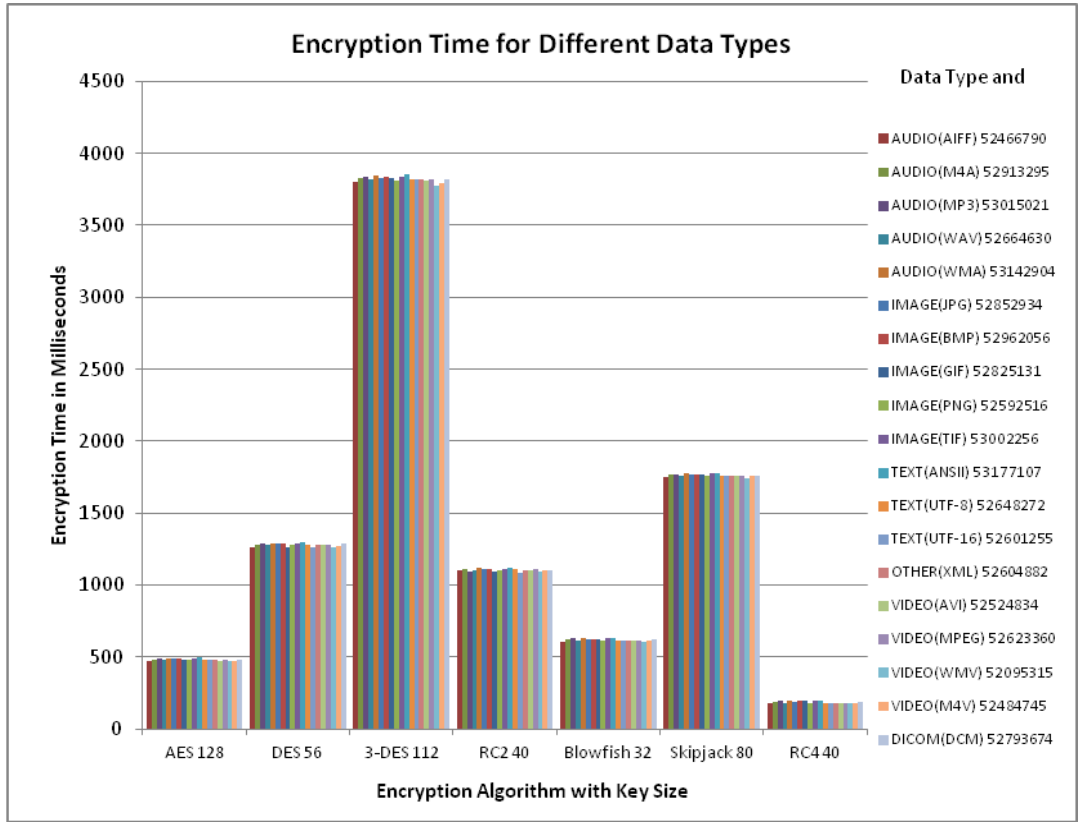

Figure 1. Encryption time Vs Cipher Algorithm for files of different data type

**Observation:** The result shows that the encryption time does not vary according to the type of the data. Encryption depends only on the number of bytes in the file and not on the type of file. AES works faster than other block ciphers. RC4 with key size 40 is fastest among the algorithms tested.

**Case Study 2: Data files of same type with different sizes.**

This case study is taken to ensure once again the observations obtained in case study 1. Case study 1 revealed that encryption time depends on number of bytes in the file. To ensure this another study is made in which different files of same types but different sizes are given for encryption and estimated the encryption time. For all executions key size and block mode are kept at bare minimal parameters. Table 2 gives the details about the files used for all executions and Fig. 2 and 3 show the execution results.

Table 2. Execution Parameters for files of different size.

| File Type | Varying Parameters (Data Size) | Constant Parameters |
|---|---|---|
| AIFF | 10.7MB, 50MB, 100MB | Data Type, Key size |
| AVI | 50MB, 100MB, 482MB | |

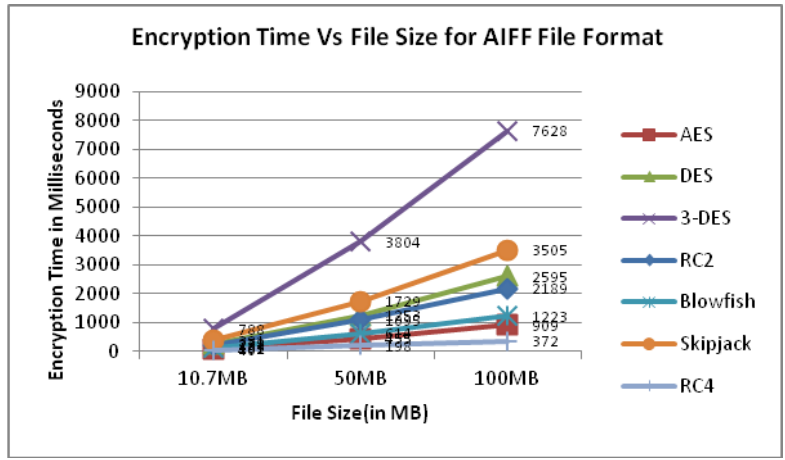

Figure 2. File size Vs Encryption time for AIFF file of different sizes.

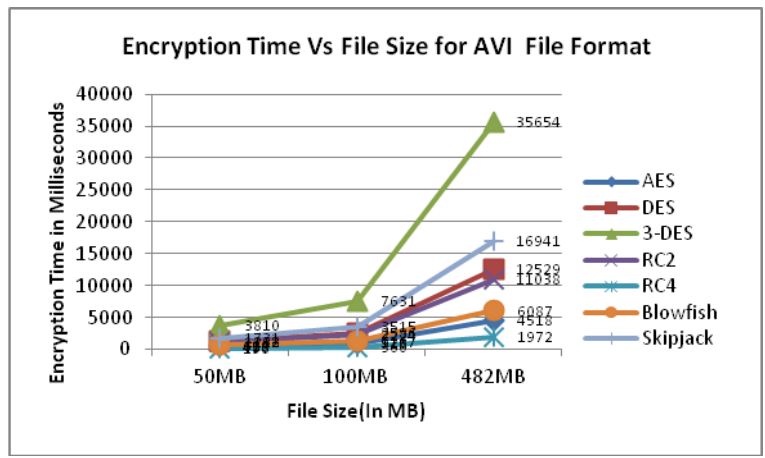

Figure 3. File size Vs Encryption time for AVI file of different sizes.

Table 3. Encryption time for files of different sizes

| File Type | Size (In MB) | Encryption Time in Millisecond | | | | | | |
|---|---|---|---|---|---|---|---|---|
| | | AES | DES | 3-DES | RC2 | Blowfish | Skipjack | RC4 |
| | | 128 | 56 | 112 | 40 | 32 | 80 | 40 |
| AIFF | 10.7 | 101 | 272 | 788 | 238 | 133 | 381 | 40 |
| | 50 | 455 | 1253 | 3804 | 1095 | 614 | 1729 | 198 |
| | 100 | 909 | 2595 | 7628 | 2189 | 1223 | 3505 | 372 |
| AVI | 50 | 456 | 1268 | 3810 | 1112 | 629 | 1731 | 196 |
| | 100 | 918 | 2586 | 7631 | 2224 | 1267 | 3515 | 360 |
| | 482 | 4518 | 12529 | 35654 | 11038 | 6087 | 16941 | 1972 |

**Observation:** From the results in Table 3 and Fig. 2 and 3 we can find that the result for different size of data varies proportional to the size of data file. Encryption time increases as file size increases in multiples of data size. For each encryption algorithm same parameters are used for files of different sizes.

### Case Study 3: File with different data densities.

This case study is taken to check whether the encryption depends on density of data or not. Encryption rate is evaluated for the two different data density file; a sparse file of 69MB and a dense file of 58.5MB. For a cipher algorithm, key size and block mode are kept at bare minimal parameters. The results of execution are shown in Table 4.

Table 4. Execution rate for sparse and dense data file

| Algorithm Name | Sparse (72000118 Bytes) AIFF file | | Dense (61392454 Bytes) AIFF file | |
|---|---|---|---|---|
| | Encrypt Time(ms) | Encryption Rate(MB/s) | Encrypt Time(ms) | Encryption Rate(MB/s) |
| AES 128 | 634 | 108.28 | 540 | 108.40 |
| DES 56 | 1801 | 38.11 | 1537 | 38.08 |
| 3-DES 112 | 5076 | 13.52 | 4365 | 13.41 |
| RC2 128 | 1520 | 45.16 | 1285 | 45.55 |
| Blowfish 128 | 854 | 80.38 | 723 | 80.96 |
| Skipjack 128 | 2386 | 28.77 | 2042 | 28.66 |
| RC4 128 | 253 | 271.35 | 216 | 271.01 |

**Observation:** Encryption rate for sparse and dense file has been calculated. The Table 4 shows that the encryption time is not affected by density of data in a file. The variation in time with respect to different algorithms follows the same pattern for both sparse and dense files. The encryption rate for a particular cipher algorithm remains the same, even if the file is sparse or dense. It depends on only the number of bytes in the file.

### Case Study 4: Encryption Algorithms with different key sizes

This case study is to analyze the effect of changing the size of encryption key on encryption time. BMP file of 50.5MB is taken and different cipher algorithms are executed for different size of keys supported by them in ECB mode with PKCS#5 padding scheme. The various key sizes mentioned in Table 1 are used during experimentation. Fig. 4 shows the result of execution for key size variation.

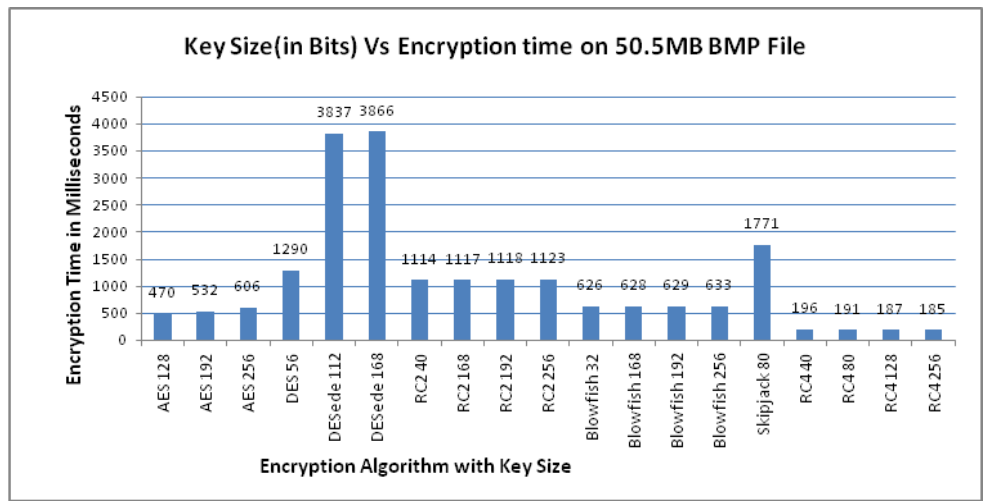

Figure 4. Variation of key sizes for different cipher Algorithms

**Obsevation:** The execution results show that for all ciphers algorithms, the encryption time varies with the change in the size of the of the key. Encryption time increases with increase in key size for block ciphers. The variation in time is very small. AES dominates in the block cipher. RC4 is fastest among all algorithms tested.

## 6. CONCLUSION

In this paper different symmetric key algorithm have been analyzed for various parameters like different data type, data size, data density, key size, cipher block modes and tested how the encryption time varies for different algorithms. From the execution results it is concluded that encryption time is independent of data type and date density. The research shown that, encryption time only depends upon the number of bytes of the file. It also reveled that encryption time varies proportionally according to the size of data. For all block cipher algorithms that are analyzed, with increase in key size, encryption time also increases, but reduces with increase in key size for RC4. AES is fastest block cipher, but RC4 appears to be fastest among all analyzed ciphers.

## ACKNOWLEDGEMENT

We would like to thanks C-DAC, Pune for all their support during the project.

## Authors

**Ranjeet Masram** is M tech Student in Computer Engineering from College of Engineering, Pune (India). He is appointed as JRF for Joint project on Medical data Security between C-DAC,Pune and College of Engineering, Pune for a period of one year.

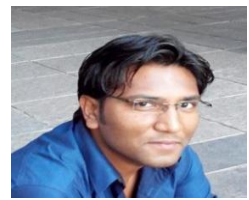

**Vivek Shahare** received his Bachelor Degree in Computer Science and Engineering from Government College of Engineering, Amravati(India). He is appointed as JRF for Joint project on Medical data Security between C-DAC, Pune and College of Engineering, Pune for a period of one year.

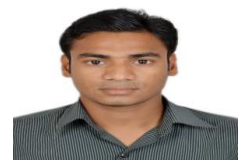

**Dr. Jibi Abraham** is Professor at College of Engineering, Pune. She received her Doctor of Philosophy (PhD) in Computer Engineering from Visvesvaraya Technological University. She is the Principal Investigator from COEP for Joint project on Medical data Security between C-DAC, Pune and College of Engineering.

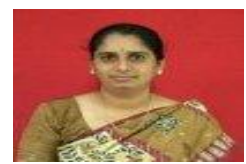

**Dr. Rajni Moona** was project engineer at IIT Kanpur. She was Visiting faculty at International Institute of Information Technology. She is the co-investigator for Joint project on Medical data Security between C-DAC, Pune and College of Engineering.

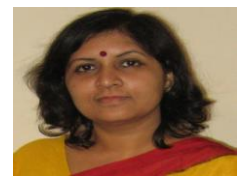

**Dr.Pradeep K.Sinha** is Senior Director, C-DAC,Pune. Dr. P.K. Sinha, Programme Coordinator, High Performance Computing and Communication (HPCC) Group, was included in the Sixteenth Edition of the "**MARQUIS Who's who in the World**," which is a prestigious international registry of outstanding men and women in a wide range of professions and geographical locations in the world. He is a visiting faculty at college of Engineering Pune. He is the first Indian to be conferred the Distinguished Engineer '09 honour.

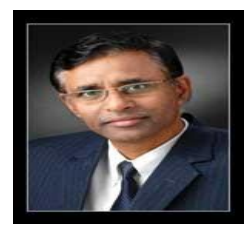

**Mr. Gaur Sunder** is the Coordinator and Head of the Medical Informatics Group (MIG) at C-DAC, Pune. He is Computer Scientist working in HPC, Distributed Systems, Cloud Computing & Virtualization, Cluster & Grid Computing, Data Repository (Big-data), Imaging, Networking, Platform (Win/Lin) and Web Technologies, and allied areas.

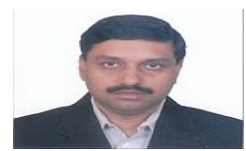

**Mr. Prashant Bendale** is Senior Technical Officer at C-DAC, Pune. His research areas include medical informatics standards, distributed / cloud computing technologies. He led the software development kits activities for medical informatics standard like DICOM & HL7.

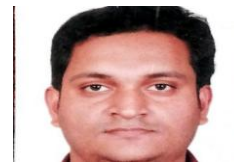

**Mrs. Sayali Pophalkar** is a project Engineer at C-DAC, Pune.